\documentclass[aps,prl,twocolumn]{revtex4}  
\usepackage[utf8]{inputenc}
\usepackage[T1]{fontenc}
\usepackage{color}
\usepackage{epsfig,amsfonts}
\usepackage{graphicx}
\usepackage{float} 
\usepackage{comment} 
\usepackage{epstopdf}
\usepackage{blindtext/blindtext}
\usepackage{bbold/bbold}

\def\be{\begin{equation}}
\def\ee{\end{equation}}

\def\bea{\begin{eqnarray}}
\def\eea{\end{eqnarray}}

\def\ben{\begin{enumerate}}
\def\een{\end{enumerate}}

\def\bea{\begin{eqnarray}}
\def\eea{\end{eqnarray}}

\begin{document}

\title{Strong-coupling emergence of dark states in XX central spin models }
\author{Claude Dimo${}^{1,2,3}$, Alexandre Faribault${}^{1}$}
\affiliation{$^1$Universit\'e de Lorraine, CNRS, LPCT, F-54000 Nancy, France}
\affiliation{$^2$Physics Department and Research Center OPTIMAS, Technische Universit\"{a}t Kaiserslautern, 67663 Kaiserslautern, Germany}
\affiliation{$^3$Universit\'e de Dschang, UR-MACET, 96 Dschang, Cameroon}
\begin{abstract}
It was recently shown that the XX central spin model is integrable in the presence of a magnetic field perpendicular to the plane in which the coupling exists. A large number of its eigenstates are such that the central spin is not correlated to the environmental spins it is coupled to. In this work, we first demonstrate that the XX-central spin model remains integrable in the presence of an arbitrarily oriented magnetic field. We then show that, provided the coupling is strong enough, dark states can actually be found even in the presence of an in-plane magnetic field. We finally provide a simple explanation of this result and demonstrate its universality for a variety of distinct distributions of the coupling of the central spin to the various bath spins.

\end{abstract}

\maketitle

\section{Introduction}

The integrability of the XX central spin model in the presence z-axis magnetic field :
\bea
\hat{H} = B^z_0\hat{S}^z_0 +  \sum^{N-1}_{k = 1} \Gamma_k \left(\hat{S}^x_0\hat{S}^x_k + \hat{S}^y_0\hat{S}^y_k\right)
\label{XXU1}
\eea
has recently been demonstrated. Moreover, it was shown that a fraction of its eigenstates can be characterised as dark states for which the central spin $\hat{S}_0$ remains, independently of the magnitude of the coupling to the $k=1 \dots N-1$ bath spins, in eigenstates $\left|\uparrow\right>$ or $\left|\downarrow\right>$ of the magnetic field part of the hamiltonian: $B_0^z \hat{S}^z_0$, as if completely decoupled from the bath spins \cite{XXdark}. The idea was further studied by explicitly showing the existence of these dark states through a Bethe Ansatz approach \cite{XXBethe}. The dark state structure was also shown to be robust against certain perturbations \cite{XXperturb}.
In spin qubits, based on the spin of single electrons trapped in quantum dots \cite{loss}, the coupling of the qubit to the bath of environmental spins is detrimental in that it ultimately leads to decoherence of the central spin and to the loss of the quantum information it should store \cite{deco1,deco2,deco3}. Dark states (and dark subspaces of the Hilbert space) then become remarkably desirable since they  can provide protection against these bath-induced negative effects leading to long lived quantum states of the qubit \cite{control1,control2} both in nitrogen-vacancy centers in diamond \cite{laraouinv,dobrovnv,hallnv} and in semiconductor quantum dots \cite{ramsaysemi,hansonsemi,schliemannsemi}.

\section{Integrability of XX models in a generic magnetic field}

In this work we study the fate of these dark states in XX central spin models submitted to an arbitrarily oriented magnetic field. Adding XY-plane components of the magnetic field to (\ref{XXU1}), breaks the rotational U(1)-symmetry of the model and therefore, the total z-axis magnetisation of the system is no longer conserved, and the eigenstates no longer have a fixed $\sum_{i=0}^{N-1} S^z_i$. Nonetheless, they remain integrable, a fact that we now demonstrate by taking an appropriate limit of the integrable $N$ spin-1/2 non-skew symmetric elliptic Richardson-Gaudin (RG) models. The latter are defined by a set of $N$ operators \cite{skrypnyk,dimoclaeys}:

\noindent
\begin{eqnarray}
\hat{R}_j
&=&
\sum^{N-1}_{k =0 (\neq j)}\left(\Gamma^x_{jk}\hat{S}^x_j\hat{S}^x_k + \Gamma^y_{jk}\hat{S}^y_j\hat{S}^y_k
+ \Gamma^z_{jk}\hat{S}^z_j\hat{S}^z_k \right)
\nonumber\\ &&+ B^x_j\hat{S}^x_j
+ B^y_j\hat{S}^y_j
+ B^z_j\hat{S}^z_j,
\label{conserved}
\end{eqnarray} 

\noindent where, in any given direction $\alpha \in \{x,y,z\}$, the coupling constants and magnetic field components, are given by:

\bea
\Gamma_{jk}^\alpha = g\frac{\sqrt{\left(\epsilon_j + j_\alpha\right)\left(\epsilon_k + j_\beta\right)\left(\epsilon_k + j_\gamma\right)}}{\epsilon_k - \epsilon_j}, \ B^\alpha_j = \frac{B^\alpha}{\sqrt{\epsilon_j + j_\alpha}} 
\label{param}
\eea

\noindent with $\beta$ and $\gamma$ the two directions ($\ne \alpha$). These models are known to be integrable in that, for arbitrary values of the parameters $\epsilon_a$, $j_\alpha$ and $B^{\alpha}$, the $N$ operators (\ref{conserved}) commute with one another and consequently share a common set of eigenstates.

We first choose $j_y = j_x$ leading to an XXZ-model where the couplings along both the x and y directions are equal. Taking now the specific value $\epsilon_0 = - j_z + \Delta$, choosing $B^z \equiv B_0^z \sqrt{\Delta} $, and finally taking the $\Delta \to 0$ limit of the operator $\hat{R}_0$, one then finds that the XX-model, in an arbitrarily oriented magnetic field: 
\bea \hat{H} \equiv \lim_{\Delta \to 0} \hat{R}_0&=& B_0^z \hat{S}^z_0 
+ B_0^x \hat{S}^x_0
+ B_0^y \hat{S}^y_0 
\nonumber\\&& 
+ \sum^{N-1}_{k = 1} \Gamma_k \left(\hat{S}^x_0\hat{S}^x_k  +\hat{S}^y_0\hat{S}^y_k\right)
\label{firstform}
 \eea
\noindent is also integrable. It remains so for arbitrary components of the magnetic field since $B_0^z$, $B_0^x \equiv \frac{B^x}{\sqrt{j_x - j_z}}$ and $B_0^y \equiv \frac{B^y}{\sqrt{j_x - j_z}}$ can all be chosen freely, independently of one another. The $N-1$ values of the resulting couplings in $\hat{H}$ are given by $\Gamma_k \equiv  g \frac{\sqrt{\left(j_x - j_z\right)\left(\epsilon_k+ j_x\right)\left(\epsilon_k + j_z\right)}}{\epsilon_k + j_z} $ and can therefore all be arbitrarily chosen while maintaining integrability using the $N-1$ free parameters $\epsilon_k$. By taking the same limit for the other $\hat{R}_{j>0}$ operators, one directly shows that the hamiltonian (\ref{firstform}) commutes with the $N$ following conserved charges:
\bea
\hat{R}_j &&= \frac{B^x}{\sqrt{\epsilon_j + j_x}}\hat{S}^x_j + \frac{B^y}{\sqrt{\epsilon_j + j_x}}\hat{S}^y_j + \sum_{\alpha}^{x,y,z}\sum_{k\neq 0,j}^{N-1}\Gamma_{jk}^{\alpha}\hat{S}^\alpha_j\hat{S}^\alpha_k
\nonumber\\&&
-g\frac{\sqrt{\left(j_x - j_z \right)\left(j_x - j_z\right)\left(\epsilon_j + j_z\right)}}{\epsilon_j  + j_z}\hat{S}_0^z\hat{S}_j^z,
\eea
\noindent with the values of $\Gamma^\alpha_{jk}$ given by eq. (\ref{param}) and $j_y = j_x \equiv j_\perp$.

In order to access the spectrum and the expectation values of local spin operators for each eigenstate of the system, we rely on the fact that the RG conserved charges (\ref{conserved}) obey simple quadratic relations so that for an eigenstate labelled $\left| \psi_n\right>$, one can find the set of $N$ eigenvalues $r_j^{n}$ such that: $\hat{R}_j \left| \psi_n\right> = r_j^{n} \left| \psi_n\right>$ by simply solving quadratic equations \cite{dimoquad,skrypnykquad}. Each ensemble of eigenvalues $(r_0^{N-1} \dots r^{n}_{N-1})$ defining the eigenstate $n$, corresponds to one of the solutions of the system:
\bea
\left[r_j^{n}\right]^2 = -\frac{1}{2}\sum_{k \neq j}C_{jk} r_k^{n} + \sum_{\alpha}\sum_{k \neq j}\left[\frac{\Gamma_{jk}^{\alpha}}{4}\right]^2 + \sum_{\alpha}\left[\frac{B_j^{\alpha}}{2}\right]^2,
\label{quadeq}
\eea
\noindent with $C_{jk}= - g \displaystyle \frac{\sqrt{\left(\epsilon_k+j_\perp\right)^2\left(\epsilon_k+j_z\right)}}{\epsilon_k-\epsilon_j}$, with $\epsilon_0 = -j_z$ so that $C_{j0}= 0$. 

Specific solutions to this system can be found numerically by smoothly deforming a given $g=0$ solution to desired coupling amplitude $g$ \cite{faribaultalgo}. Those $g=0$  solutions are simply the various tensor products $\bigotimes_{i=0}^{N-1}\left|\pm_i\right>$ built from the eigenstates $ \left|\pm_i\right>$ of $\hat{R}_{i}^{g=0} = \vec{B}_i \cdot \vec{\hat{S}}_i$. This allows us to index individual eigenstates, at finite coupling, simply by specifying the $g=0$ parent state which deforms into it. These will be denoted by the sequence of $g=0$ eigenstates, ordered by spin index from left to right. For example, $\left| + - - + \dots \right>$ means the central spin $S_0$  is in the eigenstate of $\vec{B}_0 \cdot \vec{\hat{S}}_0$ with eigenvalue $+\frac{1}{2}$, spin $S_1$ in the eigenstate of $\vec{B}_1 \cdot \vec{\hat{S}}_1$ with eigenvalue $-\frac{1}{2}$ , and so on.

The Hellmann-Feynman theorem gives the expectation values $\left<\psi_n\right| \hat{S}^\alpha_j \left|\psi_n \right>$ in terms of derivatives of the eigenvalues with respect to the proper hamiltonian's parameter. In the specific problems of interest here, one has:
\bea
\left< \hat{S}^{x,y}_j\right> &=& \sqrt{\epsilon_j + j_\perp } \ \frac{\partial{r^{n}_j}}{\partial{B^{x,y}}} \ \mathrm{\ \ with \  \ } \epsilon_0 = -j_z
\nonumber\\ \left< \hat{S}^z_0\right> &=&  \frac{\partial{r^{n}_0}}{\partial{B^z_0}};
\ \left< \hat{S}^z_{j \ne 0}\right> = \sqrt{\epsilon_j + j_z}\ \frac{\partial}{\partial{B^z_0}} \left[\frac{r^{n}_j}{\sqrt{\Delta}}\right]
\eea
Once the values of $r^{n}_j$, defining an eigenstate, have been found, these derivatives are accessible by solving the linear system of equations obtained after taking the derivatives of (\ref{quadeq}) with respect to the parameter of interest \cite{dimoclaeys}.

\section{Dark states}

As mentioned in the introduction, dark states are known occur in the XX central spin model subjected to a z-oriented magnetic field. They are characterised by the central spin being, for any coupling amplitude, exactly in one of the two eigenstates of the $B_z \hat{S}_0^z$ operator \cite{XXdark}. They can therefore, in general, be written as a tensor product $\left| \uparrow_0 \right> \otimes \left| \psi^{\mathrm{bath}}_n\right>$ $\left( \mathrm{or} \left| \downarrow_0 \right> \otimes \left| \psi^{\mathrm{bath}}_m\right>\right)$ where the various possible states of the bath spins can be found as solution to a set of reduced Bethe equations \cite{XXBethe}. 

If this product state structure can be maintained in the presence of in-plane components of the magnetic field, it could possibly lead to a dark state in which the central spin would be in a well defined eigenstate of the arbitrarily oriented $\vec{B}_0 \cdot \vec{\hat{S}}_0$. On the other hand, the mechanism through which dark states arise could require that the magnetic field is orthogonal to  the XY-plane in which the coupling occurs. In this case, an XY-plane component of the magnetic field would be sufficient to correlate the central spin and the bath preventing the appearance of dark states.

One can work with a  definition of a dark state, valid for an arbitrarily oriented central spin, by simply requiring a tensor product structure, making the reduced density matrix of the central spin describe an arbitrary pure state. On the other hand, a generic (bright) eigenstate of this coupled system would, typically, lead to a reduced density matrix for which the central spin is in a statistically mixed state. Differentiating between dark an bright states is then possible by simply computing the quantity $\gamma_0 = \left<\hat{S}_0^x\right>^2+\left<\hat{S}_0^y\right>^2+\left<\hat{S}_0^z\right>^2$, we will dub the purity factor. Indeed, $\gamma_0 = \frac{1}{4}$ for any central spin pure state: $\alpha \left|\uparrow_0\right>+\beta \left|\downarrow_0\right>$ and systematically gets reduced  ($\gamma_0 < \frac{1}{4}$) when the central spin is entangled with the bath and can only be described by a mixed state. This generalizes the simple definition $\left| \left<S^z_0\right> \right|= \frac{1}{2}$ used, for example, in \cite{XXperturb} and will provide an alternative to the entanglement entropy which is easier to compute.

\begin{figure}[h!]
\includegraphics[width=\columnwidth]{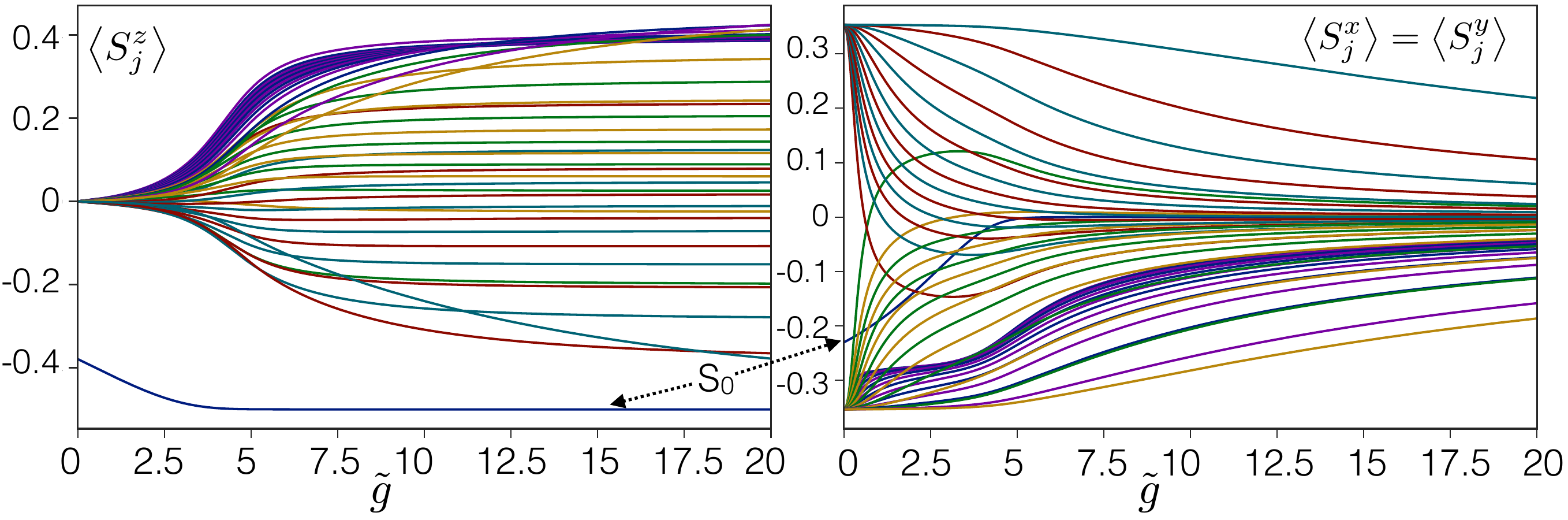}
\caption{Expectation values of the $N=42$ spins in the dark state resulting from the deformation of the $g=0$ state $ \left|- - + - - + \dots \right>$. The coupling constants are given by $\Gamma_j = 7.254 \sqrt{N-j}$ for $j =1 \dots N-1$, the magnetic field is oriented at azimuthal angle $\theta \approx \frac{\pi}{4}$ ($B^x_0 = B^y_0 = 2.23; \ B^z_0 = 3.162)$. }
\label{darkall}
\end{figure}

In Fig. \ref{darkall} we first present the expectation values of the z and the in-plane $x$ and $y$ (both identical by symmetry since the external field is chosen as $B_x = B_y$) in the eigenstate whose $g=0$ parent state is given by an alternance of two negative and one positive eigenvalues: $\left| --+--+--+ \dots\right>$. They are presented, for every spin in the system, as a function of the rescaled coupling $\tilde{g} =  \displaystyle\frac{1}{|B|}\sum_{j=1}^{N-1} \Gamma_j$ which characterizes the ratio between the total coupling to the bath and the local magnetic field term of the XX hamiltonian (\ref{firstform}).

 At $g=0$, the central spin $S_0$ is in the pure state which is the eigenstate of $\vec{B}_0 \cdot \vec{\hat{S}}_0$ with eigenvalue $-\frac{1}{2}$. As the coupling is increased, we see that if finally reaches the down pointing state (eigenstate $B_z S^z_0$ with eigenvalue $-\frac{1}{2}$) at large enough coupling. This resulting strong coupling state is therefore also a pure state of the central spin, oriented just like the dark states found for the U(1)-symmetric case (\ref{XXU1}) with a z-oriented magnetic field.

The bath spins, at g=0, all start as $\pm$ eigenstates of the $\vec{B}_j \cdot \vec{\hat{S}}_j$ which are polarized in the XY-plane, since the magnetic fields $\vec{B}_j$ has $B^z_j =0$ (see eq. \ref{conserved}). As the coupling gets stronger,  we see in the same figure that they get tilted out of that plane and gain an important z-component expectation value. The central spin has reached its strong coupling z-polarized state at $\tilde{g} \approx 5$ for the parameters used in this calculation. However, the bath spins evolve much more slowly with coupling since, at the same value of $\tilde{g}$, they are still significantly changing with increasing coupling. In the end, at strong enough coupling,  their XY-plane component finally decreases as $\displaystyle\frac{1}{g}$, as we will see in the upcoming discussion. 

One should notice that, at $\tilde{g} \approx 5$, the $B^x_0,B^y_0$ components of the magnetic field  are, in no way, negligible when compared to the total coupling to the bath. Nonetheless, the central spin appears to already reproduce the dark state behavior expected in the absence of those in-plane components of the external field.

\begin{figure}[h!]
\includegraphics[width=\columnwidth]{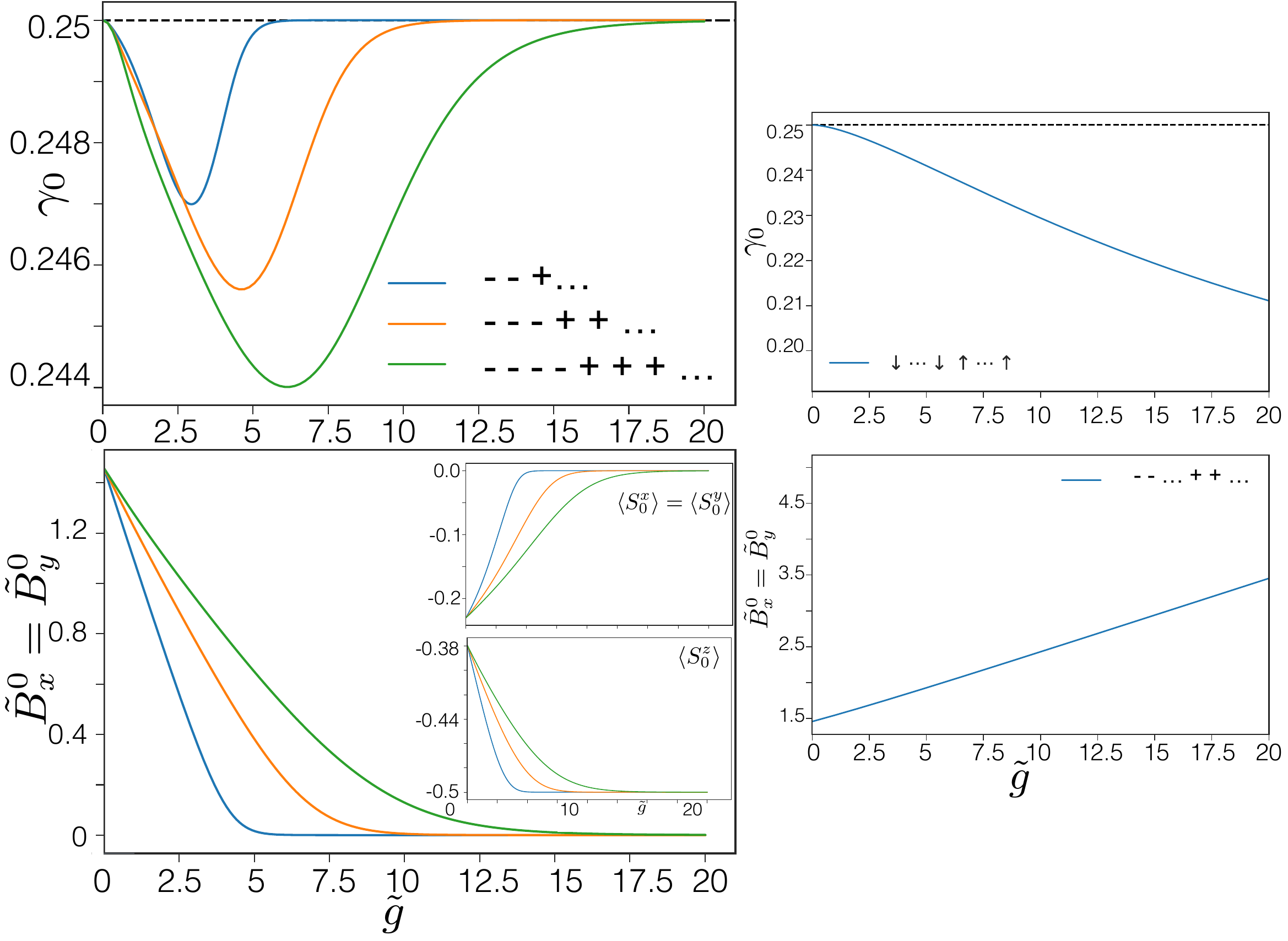}
\caption{The purity factor $\gamma_0$ of the central spin (top left panel) and the effective in-plane magnetic field acting on the central spin (bottom left panel) in 3 dark states ($g=0$ parent states given by: $\left| - - + - - + - - + \dots\right>$, $\left| - -  - + + - -  - + +\dots\right>$ and $\left| - -  - - + + + - -  - - + + +\dots\right>$) . Insets: Expectation values of the central spin' components. Smaller figures on the right show a bright state}.
\label{darkbright}
\end{figure}

In order to better understand the physics at play we present, in Fig. \ref{darkbright}, the $\gamma_0$ purity factor for four different eigenstates using the same set of parameters as for Fig. \ref{darkall}. Three of them (including the state detailed in Fig. \ref{darkall}) are presented in the top left panel and form a dark state at strong coupling. The fourth state, shown in the smaller figures on the right, remains a bright state for all finite coupling as demonstrated by the fact that $\gamma_0< \frac{1}{4}$. 

One notice that the states which ultimately become dark states at strong coupling, see their "darkness" reduced in the intermediate coupling regime. The central spin is then in a statistically mixed state; not in a pure state on the Bloch sphere. Adding an in-plane magnetic field to (\ref{XXU1}) suppresses the dark state structure over a range of relatively weak couplings. Nonetheless, dark states reemerge at strong enough coupling although, as noted before, they do so even when the in-plane fields are not negligible before the coupling term. This seems to point to the existence a mechanism which allows the central spin not to feel the in-plane magnetic field. A clear understanding of the process involved in this reemergence of dark states can actually be obtained.

Plotted in the middle panel, right below the purity factor of the three reemerging dark states, is the effective mean-field in-plane magnetic field felt by the central spin when considering the Overhauser field contribution of the bath spins. By replacing, in a typical mean-field approach, the bath spins by their expectation values, one can indeed build an effective hamiltonian for the central spin which is simply given by:
\bea
H_{\mathrm{eff}} &=& \vec{B}_0 \cdot \vec{\hat{S}}_0 
+ g\sum^{N-1}_{k = 1} \tilde{\Gamma}_k \left[\hat{S}^x_0 \left<\hat{S}^x_k \right> +\hat{S}^y_0\left< \hat{S}^y_k \right>\right]
\nonumber\\&=& 
\vec{\tilde{B}} \cdot \vec{\hat{S}}_0,
\eea
\noindent where the various effective magnetic field components are given by: 
\bea
\tilde{B}^{x,y} &=& B^\alpha_0+g\displaystyle \sum^{N-1}_{k = 1} \tilde{\Gamma}_k  \left<\hat{S}^{x,y}_k \right>; \  \ \
\tilde{B}^z = B^z_0. 
\label{effectivefield}
\eea
\noindent having defined $\Gamma_k \equiv g \tilde{\Gamma}_k$. Judging from these figures, one can clearly see that the return to a pure state ($\gamma_0 = 0.25$) is perfectly correlated with the coupling strength at which the effective in-plane components of the field reach $\tilde{B}_x=\tilde{B}_y =0$. As was seen in Fig. \ref{darkall}, at strong enough coupling, one can reach a regime where the expectation value of the in-plane components of the bath spins has become linear in $\frac{1}{g}$, allowing the cancellation of the effective in-plane magnetic field to maintain itself over the whole range of large couplings. This cancellation can only occur for couplings large enough to have: $\left|\sqrt{(B^x_0)^2+(B^y_0)^2 }\right| \displaystyle \le \frac{1}{2}\left| \sum^{N-1}_{k = 1} \Gamma_k \right|$, a limit imposed by the fact that the norm of the expectation value of a spin in any given direction has to be  $\le \frac{1}{2}$. This is a much less stringent requirement on the coupling strength than $\displaystyle\frac{\left|\sqrt{(B^x_0)^2+(B^y_0)^2 }\right|}{\left| \sum^{N-1}_{k = 1} \Gamma_k \right|} \ll 1$ would be. The effective cancellation of the in-plane field by the Overhauser contribution of the bath therefore explains why dark states can reemerge despite the fact that the in-plane magnetic field is not yet negligible in front of the coupling strenght.

\begin{figure*}[t]
\includegraphics[width=0.65\columnwidth]{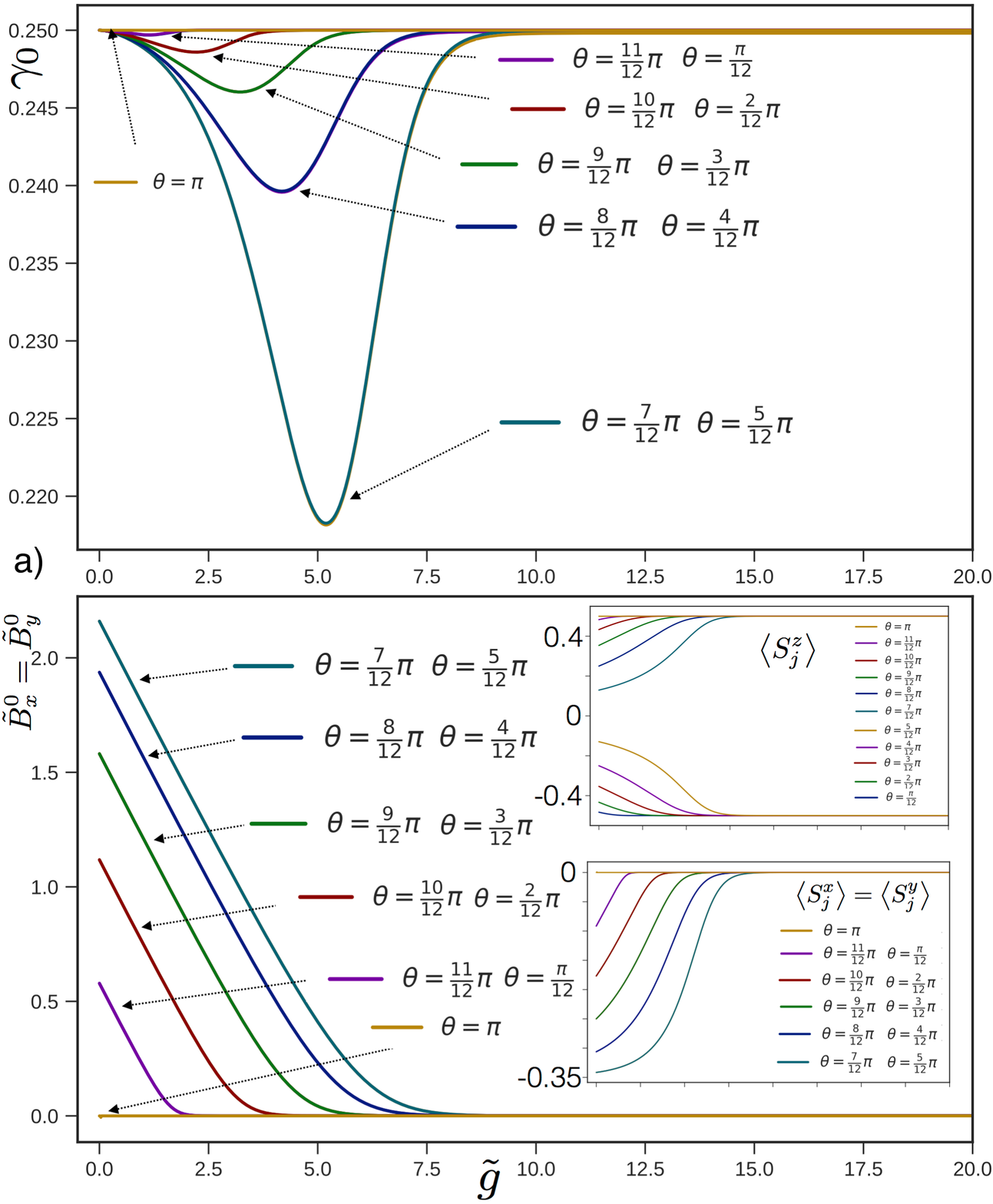}
\includegraphics[width=0.65\columnwidth]{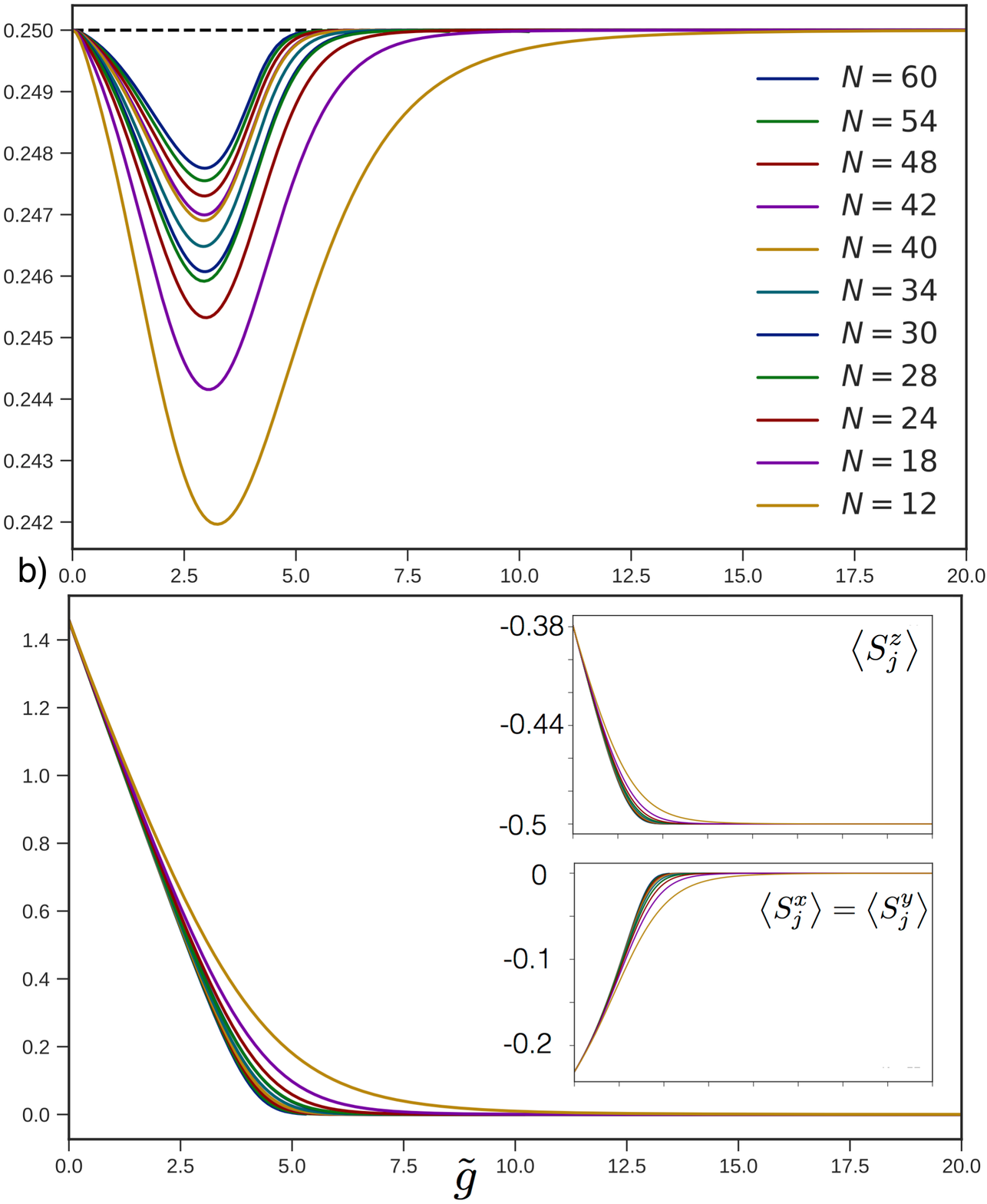}
\includegraphics[width=0.65\columnwidth]{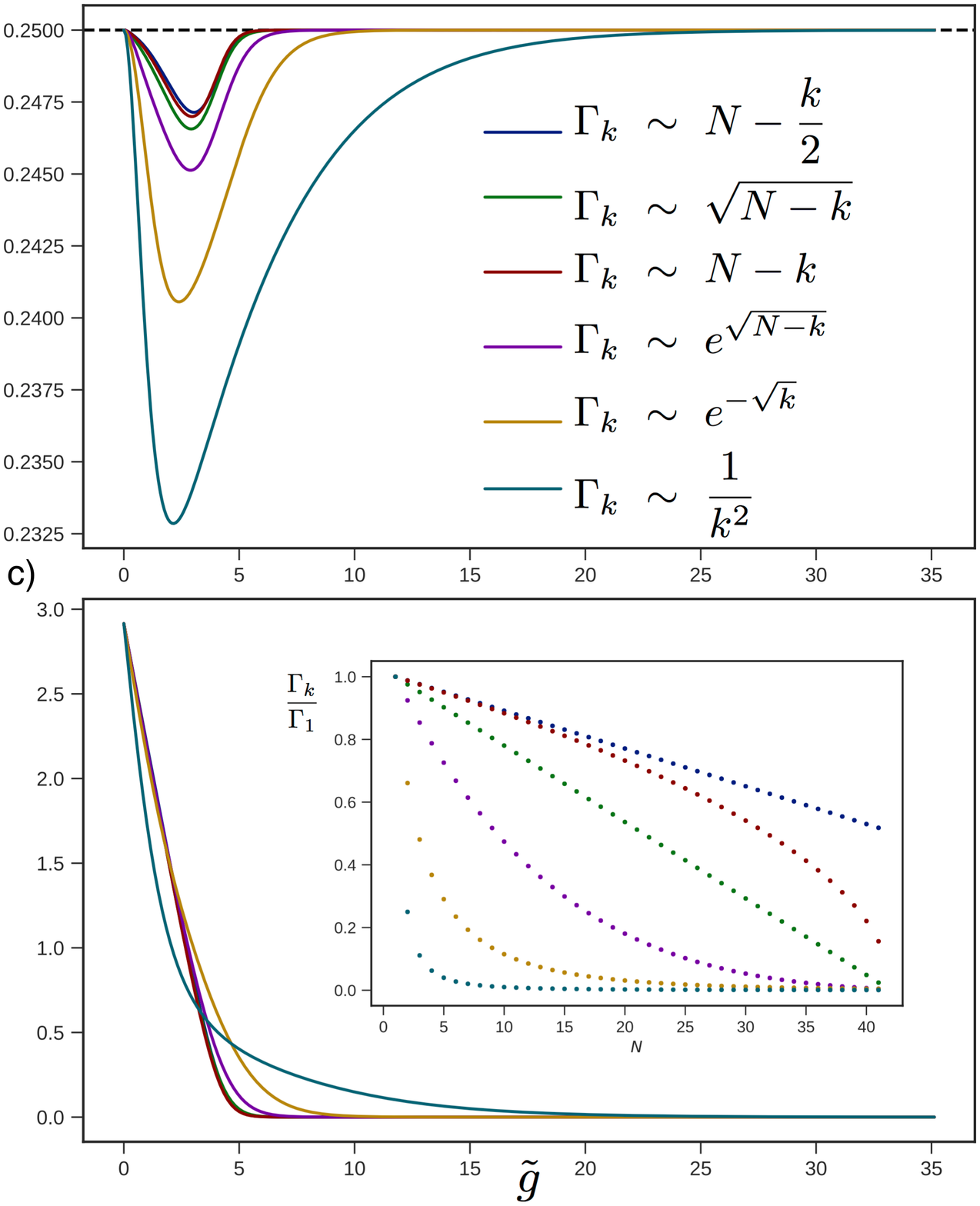}
\vspace{-1cm} \caption{ {\bf Panel a)} Top:The purity factor $\gamma_0$ of the central spin in a dark state, as a function of the rescaled coupling strenght for a variety of magnetic field orientations $B^z_0 = |B| \cos(\theta),B^x_0=B^y_0 =  \frac{|B|}{\sqrt{2}} \sin(\theta)$, keeping a fixed norm $|B|$. Bottom: Effective in plane magnetic field acting on the central spin: eq. (\ref{effectivefield}). Parameters are the same as in Fig. \ref{darkall}. {\bf  Panel b)} Purity factor (Top) and effective in-plane field (Bottom) for a variety of system size. Coupling constants are given by $\Gamma_k = C_N \sqrt{N-k}$ with the $C_N$ constants chosen to ensure that $\sum_{k=1}^{N-1}\Gamma_k$ is the same for all sizes $N$.
{\bf  Panel c)} Purity factor (Top) and effective in-plane field (Bottom) for different coupling distributions (always with $N=42$ spins) sharing the same value of the total coupling $\sum_{k=1}^{N-1} \Gamma_k$. Inset: Distribution of coupling constants.}
\label{alldata}
\end{figure*}

For a fixed magnitude of the field $\vec{B}_0$ we present, in Fig. \ref{alldata} a), the purity factor and effective magnetic field, for different values of the external field's azimuthal angle $\theta$ . One sees that the range of coupling strength over which the purity factor of the central spin state is lower than $\frac{1}{4}$ becomes wider with increasing angle. These results are completely consistent with the physical picture proposed in this work, namely that dark states reemerge, at large enough coupling, when the Overhauser field cancels the in-plane applied field.  In the weak coupling regime where dark states are suppressed, the deeper dip in the purity factor of the central spin indicates that the entanglement between the central spin and the bath gets larger for more strongly titled magnetic fields. Since the total magnetic field has the same magnitude for each of the curves, one clearly understands that the underlying physics is exclusively controlled by the magnitude of the in-plane components of the magnetic field (responsible for breaking the U(1)-rotational symmetry around the z-axis of hamiltonian (\ref{XXU1})). Since the bath spins get a non-zero expectation value $\left<S^z_k\right>$, one requires stronger coupling than the minimal criterion discussed above for cancellation: $\tilde{g} \ge 2 \sin(\theta)$, which could be sufficient only if the bath spins were perfectly aligned in the XY-plane and therefore maximally contributing to the cancellation. Nonetheless, as we saw previously, dark states still reemerge at couplings considerably lower than would be required for the in plane field to become negligible. 

In Fig. \ref{alldata} b) the eigenstate which was presented in Fig. \ref{darkall} is now shown for a variety of bath sizes. The coupling constants are given by $\Gamma_k = C_N \sqrt{N-k}$ where $C_N$ are chosen in such a way that the total coupling to the bath $\sum_{k=1}^{N-1} \Gamma_k$ stays the same for every system size $N$. One finds that the dip in purity is then systematically limited to a similar range of couplings. However, the depth of this dip is reduced by increasing the number of bath spins. This allows us to infer that, in the thermodynamic limit $N \to \infty$, for this particular distribution of couplings, the dark state structure should (as is the case in the absence of in-plane magnetic field) be maintained for arbitrary values of the coupling strength $\tilde{g}$. This demonstrates that the factor controlling the purity dip is not, in itself, the total coupling to the bath but actually the way this coupling is spread out over the different spins in the bath.

We therefore present, in Fig. \ref{alldata} c) results for different coupling constant distributions, all characterised by the same total coupling and the same number of spins. These results first demonstrate that the reemergence of dark states, at strong enough coupling, is a universal feature of the model;  it happens for every distribution of couplings. However, the distribution of couplings does have an impact on the range of coupling strengths over which we deviate from a perfect dark state and the importance of this deviation. The results are highly reminiscent of the finite size scaling shown in the middle panel b) and one can conclude that the purity dip is in fact controlled by an effective number of active bath spins. Spins which are too weakly coupled to the central spin do not participate in the effective cancellation of the in-plane magnetic field by the Overhauser field. The thermodynamic limit discussed before will then depend on the nature of the coupling distribution. When the couplings decrease, with index $k$, faster than $\frac{1}{k}$, one then has $\displaystyle \lim_{N \to \infty }\sum_{k=1}^{N-1 }\frac{\Gamma_k}{\Gamma_1}$ to converge, the weakly coupled spins will not contribute. One will then still find, at $N\to \infty$, a regime at weak coupling where deviations from a pure central spin state are found. This was numerically verified explicitly for the $\frac{1}{k^2}$ distribution, for which the results stop changing significantly for $N > 14$. Indeed, for rapidly decreasing coupling constants, a larger system will only add weakly coupled spins which, as mentioned before, are unable to contribute to the Overhauser field at a given small value of $g$. In this case, the cancellation of the in-plane field cannot be achieved due to the small number of bath spins with large enough coupling to contribute. Dark states can then only appear at a larger value of $g$. On the other hand if, as in panel b), the coupling are slower than $\frac{1}{k}$ the dip will resorb in the thermodynamics limit and dark states should exist for any coupling strength.

In this work we have first shown that the XX central spin model remains integrable in the presence of XY-plane components of the magnetic field. Dark states, which are found for any coupling strength in the absence of in-plane field were then shown to get suppressed at weak couplings. However, at strong enough coupling their reemergence appears to be a universal feature of the model. It can be understood as a consequence of the restructuration of the bath spins such that the resulting mean-field Overhauser field  exactly  cancels out the in-plane magnetic field. The resulting dark states are such that the central spin becomes aligned with the $z$ direction. What we can globally conclude from this work is that, in what seems like a paradoxical statement, in the XX central spin model with in-plane field, one must increase the coupling to decouple the central spin from its environnement.

\end{document}